\documentclass[prl,twocolumn,superscriptaddress]{revtex4}
\usepackage{times, hyperref}
\usepackage{amsmath}
\usepackage{amsfonts}
\usepackage{amssymb}
\usepackage{graphicx, color}
\usepackage[normalem]{ulem}

\newcommand{\Eb}{\Delta_{\textrm{b-Exc}} }
\newcommand{\Ef}{\Delta_{\textrm{f-Exc}} }
\newcommand{\Enf}{\Delta_{\textrm{NF}} }
\newcommand{\nf}{\psi_{\textrm{NF}} }
\newcommand{\nuT}{\nu_{\textrm{T}}}

\pdfoutput=1

\begin{document}

\title{Topological exciton Fermi surfaces in two-component fractional quantized Hall insulators}

\author{Maissam Barkeshli}
\affiliation{Department of Physics, Condensed Matter Theory Center, University of Maryland, College Park, Maryland 20742, USA}
\affiliation{Joint Quantum Institute, University of Maryland, College Park, Maryland 20742, USA}
\affiliation{Kavli Institute for Theoretical Physics, University of California, Santa Barbara, California, 93106, USA}
\author{Chetan Nayak}
\affiliation{Station Q, Microsoft Research, Santa Barbara, California 93106-6105, USA}
\author{Zlatko Papic}
\affiliation{School of Physics and Astronomy, University of Leeds, Leeds, LS2 9JT, United Kingdom}
\author{Andrea Young}
\affiliation{Department of Physics, University of California, Santa Barbara, California 93106-6105 USA}
\author{Michael Zaletel}
\affiliation{Department of Physics, Princeton University, Princeton, NJ 08540, U.S.A.}
\affiliation{Station Q, Microsoft Research, Santa Barbara, California 93106-6105, USA}
\affiliation{Kavli Institute for Theoretical Physics, University of California, Santa Barbara, California, 93106, USA}


\begin{abstract}
A wide variety of two-dimensional electron systems (2DES) allow for independent control of the total and relative charge density of two-component fractional quantum Hall (FQH) states.
In particular, a recent experiment on bilayer graphene (BLG) observed a continuous transition between a compressible and incompressible phase at total filling $\nuT = \tfrac{1}{2}$ as charge is transferred between the layers, with the remarkable property that the incompressible phase has a \it finite interlayer polarizability\rm.
We argue that this occurs because the topological order of $\nuT = \tfrac{1}{2}$ systems supports a novel type of  interlayer exciton that carries Fermi statistics. If the fermionic excitons are lower in energy than the conventional bosonic excitons (i.e., electron-hole pairs), they can form an emergent neutral Fermi surface, providing a possible explanation of an incompressible yet polarizable state at $\nuT = \tfrac{1}{2}$.
We perform exact diagonalization studies which demonstrate that  fermionic excitons are indeed lower in energy than  bosonic excitons.
This suggests that a ``topological exciton metal'' hidden inside a FQH insulator may have been realized experimentally in BLG. We discuss several detection schemes by which the topological exciton metal can be experimentally probed. 
\end{abstract}

\maketitle

Two-component quantum Hall systems have long been known to host rich phase diagrams, exhibiting
intrinsically two-component fractional quantum Hall (FQH) states, broken symmetry states, and quantum
phase transitions at fixed total filling fraction $\nuT$~\cite{girvinmacdonald, eisenstein2014,shayegan1999}.
When tunneling between the components is effectively zero, the system acquires an enhanced total 
and relative $U(1)_{\textrm{T}} \times  U(1)_{\textrm{r}}$ symmetry due to the independently conserved 
charges of the two components. This situation is most easily realized when the two components are related by spin or valley symmetry~\cite{eisensteintransition1989,bishop2007,feldmantransition2013,weitz_broken-symmetry_2010,falson_even-denominator_2015}, or in double layer systems in which a barrier suppresses interlayer tunneling~\cite{eisenstein1992a,suen1992}.

A number of experimental platforms have been used to study the resulting phase diagram of two component FQH phases at $\nuT=\frac{1}{2}$,
including wide quantum wells~\cite{suen1994,liu2011,liu_even-denominator_2014,liu2014prl}, ZnO 
heterostructures~\cite{ki2014,falson_even-denominator_2015}, and, most recently, bilayer graphene (BLG), where the two components correspond to layer~\cite{zibrov2016}.
In many of these systems the relative filling $\nu_+ - \nu_-$ of the two components can be tuned \textit{in situ}.
In particular, Ref.~\onlinecite{zibrov2016} has reported the remarkable experimental observation of a BLG state at $\nuT=\frac{1}{2}$
that is incompressible, yet possesses a finite interlayer polarizability. This insulating state persists over the range of interlayer polarization  $\nu_+ - \nu_- \approx 0 - 0.18$, which is strikingly large when compared with the typical width of FQH plateaux. 
In the presence of $U(1)_{\textrm{r}}$, finite polarizability indicates a vanishing neutral gap, and hence hints at the discovery of a new phase of matter distinct from a fully gapped quantized Hall state.

In this Letter we address this experimental finding  by analyzing the possible phases which can occur in a two component system at $\nuT=\frac{1}{2}$ as density is transferred between the two components. The $U(1)_\textrm{r}$ symmetry ensures that inter-component excitons can exist as long-lived excitations.
We argue that the experimental observations of Ref.~\onlinecite{zibrov2016} could be explained by a FQH insulator whose interlayer excitons have delocalized into a degenerate quantum liquid. 

The problem is particularly rich at $\nuT=\frac{1}{2}$ because the fractionalized nature of even-denominator FQH states guarantees that in addition to the familiar bosonic exciton (b-Exc), the system also hosts a topologically non-trivial fermionic exciton (f-Exc).
If the f-Exc is lower in energy, the system naturally forms a ``topological exciton metal'' at finite f-Exc density. This new phase of matter would exhibit insulating charge transport but \emph{metallic}  counterflow resistance.

We consider two main scenarios. First we discuss systems in which the two components arise from a crossing between an $N=0$ and $N=1$  LL in the limit where the distance $d$ between them is small compared to the magnetic length $\ell_B$, as occurs in ZnO~\cite{falson_even-denominator_2015}, wide quantum wells~\cite{liu2011}, 
and BLG~\cite{zibrov2016}. Our exact diagonalization calculations show that the f-Exc is indeed lower in energy than the b-Exc. 
In the second scenario, relevant to a bilayer with $d /\ell_B \gtrsim 1$, we consider a crossing of two $N=0$ levels, 
where we also argue that the f-Exc will determine the nature of the intermediate phase.

\paragraph{$(N_+,N_-) = (1,0)$: the Pfaffian exciton metal.} In the experiment of Ref.~\onlinecite{zibrov2016}, an electric field perpendicular to the bilayer causes the first excited $N_+ = 1$ LL in the top layer to cross in energy with the lowest $N_- = 0$ LL in the bottom layer.
The filling fraction of the two layers is $\nu_+ = 1/2 - \delta$ and $\nu_- = \delta$.
Because $N_+ = 1$, when $\delta = 0$ the system is observed to form an incompressible FQH state in the top layer, roughly
analogous to the $5/2$-plateau of GaAs~\cite{willett1987}. Based on  numerical evidence~\cite{Morf98, zibrov2016}, as well as the recent experimental observation of a  half-integer thermal Hall effect~\cite{banerjee2017observation}, we will assume that the system forms a Moore-Read Pfaffian 
FQH state.\cite{moore1991} However all even-denominator FQH states must contain a charge $-e$ boson, so will lead to essentially the same conclusions.
When $\delta = 1/2$, the particles reside in an $N_- = 0$ level, so the system forms a compressible CFL~\cite{willett1997,halperin1993,jainCF}.

What is the fate of the system at intermediate $\delta$? As $\delta$ increases from zero, the top layer loses charge to 
the bottom layer. Due to the strong Coulomb interaction between layers, excitons will form, with $-e$ charge in the top layer 
and $e$ charge in the bottom layer, with a binding energy on the order of the interlayer Coulomb interaction.
Crucially, at $\nuT=\tfrac{1}{2}$ this system supports two topologically distinct types of excitons. The conventional bosonic exciton (b-Exc) is formed when an electron is transferred from the Pfaffian state in the top layer to the bottom layer. On the other hand, the Pfaffian state also has a charge $-e$ bosonic excitation, which can be thought of as a Laughlin
quasiparticle associated with inserting two flux quanta into the system. A bound state of the charge $-e$ boson in the top Pfaffian layer and an electron in the bottom is a \it fermionic \rm exciton (f-Exc).
In contrast to the b-Exc, the f-Exc is a topologically non-trivial quasiparticle; it can also be thought of as a bound state of the b-Exc and the anyonic ``neutral fermion'' $\nf$ of the Pfaffian phase.
As we will demonstrate within the long wavelength effective field theory, this f-Exc is coupled to an emergent $Z_2$ gauge field. A pair of f-Exc's is topologically equivalent to a pair of b-Exc's.

Because excitons are neutral particles, they have some non-zero dispersion $\epsilon(k)$ and can delocalize. If the excitons attract, there may be an instability and the transition will be discontinuous, but otherwise we can  consider three  types of ground states for the excitons: density-wave, condensate, and metal.\cite{Jamei}
First, depending on the interactions between the excitons, it may be preferable for the excitons to form a density-wave state, for example  stripes or a Wigner crystal.
In the presence of weak disorder that pins the density wave, this state can be viewed as a localized state of excitons, e.g. a Bose glass or Anderson insulator for the b-Exc, f-Exc respectively.\cite{FisherBoson, LeeReview}  

As the density of excitons increases with $\delta$, the b-Exc can potentially undergo a quantum phase transition to a superfluid, spontaneously breaking U(1)$_r$. Analogous to the $\nuT=1$ exciton condensate~\cite{shayegan1999, eisenstein2014}, the condensation of the b-Exc leads to an interlayer coherent Moore-Read Pfaffian state.
Alternatively, if the f-Exc are more stable, increasing their density leads to a Fermi surface whose volume is set by $\delta$.
In this case, the Pfaffian state coexists with a Fermi surface of f-Exc's,  leading to insulating charge transport but metallic counterflow.
There is no sharp transition between the Anderson insulator state and the ``metallic' state of excitons, because in two dimensions all states are localized by disorder.
At finite temperature there is a crossover from the localized to delocalized regime as the temperature is increased, with a crossover temperature $T^* \sim e^{- \epsilon_F/W}$, where $\epsilon_F$ is the Fermi energy and $W$ is the disorder strength.\cite{LeeReview}

Let us now describe the above scenario more concretely in terms of a long wavelength effective field theory.
$c_{+}$ and $c_{-}$ denote the electrons in the two layers. To describe the system at $\nuT = 1/2$, we attach
two flux quanta to each electron, to obtain composite fermions (CFs) $\psi_+$ and $\psi_-$. It is convenient to describe this in terms of
a parton construction (see e.g.~\cite{barkeshli2012hlr}) $c_\pm = b \psi_\pm$, where $b$ is a charge-$e$ boson and $\psi_+$, $\psi_-$ are the neutral CFs.
$b$ and $\psi_{\pm}$ carry charge $1$ and $-1$, respectively, under an internal emergent gauge field $a$, associated with the phase rotations $b \rightarrow e^{i\theta} b$, $\psi_{\pm} \rightarrow e^{-i\theta} \psi_{\pm}$
which keep the physical electron operator invariant. 
Introducing $A_{\textrm{T}} = A_+ + A_-$ as an external probe gauge field for $U(1)_{\textrm{T}}$, and $A_r = (A_+ - A_-)/2$ as a probe gauge-field for the $U(1)_{\textrm{r}}$, the $\psi_{\pm}$ carry charge $\pm 1/2$ under $A_r$.

Next, we assume a mean-field ansatz where $b$ forms a bosonic $\nu = 1/2$ Laughlin state, and $\langle a \rangle = 0$. The
resulting field theory can be written as
\begin{align}
\mathcal{L} = -\frac{2}{4\pi} \tilde{a} \partial \tilde{a} + \frac{1}{2\pi} (a + A_{\textrm{T}}) \partial \tilde{a} + \mathcal{L}_\psi(\psi_{\pm}, a, A_r).
\end{align}
Here $a \partial a \equiv \epsilon^{\mu\nu\lambda} a_\mu \partial_\nu a_\lambda$,
$\frac{1}{2\pi} \epsilon^{\mu\nu\lambda} \partial_\nu \tilde{a}_\lambda$ is the conserved current
for the $b$ particles, and the first term on the RHS above is the effective action
for a bosonic $1/2$ Laughlin FQH state~\cite{wen04}:
\begin{align}
\label{Lpsi}
\mathcal{L}_\psi  &= \sum_{\alpha = \pm }[\psi_{\alpha}^\dagger (i\partial_t + a_t + \alpha A_{r;t}/2) \psi_\alpha
\nonumber \\
&+ \frac{1}{2m_\alpha} \psi_\alpha^\dagger ( i \partial_i + a_i +  \alpha A_{r;i}/2)^2 \psi_\alpha + \cdots],
\end{align}
where $\cdots$ indicates higher order interactions among the CFs.
We can now consider a variety of possible mean-field states for the CFs $\psi_{\pm}$.

(1) \it Two-component composite Fermi liquid\rm. Here, $\psi_{\pm}$ both form a composite Fermi sea.
This describes a CFL state with two Fermi surfaces, with Fermi wave vectors $k_{F\pm} = \ell_B^{-1} \sqrt{2 \nu_{\pm}}$,
where $\nu_{\pm}$ is the electron filling in the two layers.
This phase is most natural when $\nu_- \sim 1/2$.

(2) \it $Z_2$ fractionalized exciton metal\rm.
We consider a state where species $\psi_+$ forms a paired state, $\langle \psi_+ \psi_+ \rangle \neq 0$,
while $\psi_-$ continues to form a Fermi surface with $k_{F-} = \ell_B^{-1} \sqrt{2 \nu_-}$. This breaks the $U(1)$ gauge symmetry
down to $Z_2$, and the Higgs mechanism sets $a +  A_r/2 = 0$. In the limit $\nu_- = 0$, we expect $\psi_+$ forms a $p_x + ip_y$ 
state since the system is described by a Moore-Read Pfaffian state in the top layer~\footnote{More generally, when $\psi_+$ forms any BCS paired
odd angular momentum state, then other variants of the Pfaffian state are realized.}~\cite{read2000}.

As $\nu_-$ is increased, the system is described by a Pfaffian state in $\psi_+$  together
with a Fermi sea of $\psi_-$.  Since we have locked $a = -A_r/2$, Eq.~\eqref{Lpsi} implies that
$\psi_-$ effectively becomes coupled only to $A_r$, with \emph{unit} charge. Physically, this implies that $\psi_-$ is a fermion which carries a unit dipole moment perpendicular to the layers, and can thus be identified with the f-Exc.

However,  $\psi_-$ is still coupled to an emergent $Z_2$ gauge field, corresponding to the remnant of $a$ after the
pairing of the $\psi_+$ fermions, reminiscent of the `orthogonal metal' phase~\cite{Nandkishore2012}.
Importantly, the $\psi_+$ and $\psi_-$ fermions are both coupled to this $Z_2$ gauge field, so are non-trivially entangled.
In particular, the f-Exc will acquire a $\pi$-phase upon encircling the Pfaffian's non-Abelian charge $e/4$ quasiparticle; hence the f-Exc will see any localized $\pm e/4$ quasiparticles pinned to the disorder potential as sources of random $\pi$-flux.

A model wave function for this state can be written as follows:
$\Psi_{\text{exFS}}(\{z_i, w_a\}) = \mathcal{P}_{\text{LLL}} \psi_{\text{f-Exc}}(\{{\bf r}_a\} ) \prod_{a<b} (w_a - w_b)^2 \prod_{i,a} (z_i - w_a)^2 \mbox{Pf}\left(\frac{1}{z_i - z_j} \right) \prod_{i < j} (z_i - z_j)^2 $.
Here $z$ and $w$ are the complex coordinates of the electrons in the top and bottom layers, respectively,
with $w_a = r_{a;x} + i  r_{a;y}$. $\psi_{\text{f-Exc}}(\{{\bf r}_a\})$ is the wave function for the excitons, which can be
taken to be in a Fermi sea. $\mathcal{P}_{\text{LLL}}$ denotes projection to the lowest LL.
While this wave function is written as if both layers are in the lowest LL,
it should be transposed to the case where the $\mbox{Pf}$ layer is in the first LL
by acting with the LL raising operator on each $z$-electron, ${\prod_i} (\partial_{z_i} - \frac{\overline{z}_i}{4\ell_B^2})$.

(3) \it Interlayer coherent FQH states: exciton condensates\rm.
We can consider a state where both $\psi_{\pm}$ CFs form a paired state, $\langle \psi_+ \psi_+ \rangle \neq 0$,
$\langle \psi_- \psi_- \rangle \neq 0$, which breaks $U(1)_{\textrm{r}}$ and gives interlayer coherence. As a result these phases have a Goldstone mode and superfluid-like counterflow. We can further distinguish two cases:

(a) $\langle \psi_+ \psi_- \rangle \neq 0$. In this case, since we also have $\langle \psi_+ \psi_+ \rangle \neq 0$, we can
treat $\langle \psi_+ \psi_- \rangle$ and $\langle \psi_+^\dagger \psi_- \rangle$ as equivalent. Since $ \psi_+^\dagger \psi_-$
carries unit $U(1)_{\textrm{r}}$ charge, its expectation value implies that the interlayer $U(1)_{\textrm{r}}$ is completely broken. This corresponds to the case where the b-Exc form a condensate.

(b)  $\langle \psi_+ \psi_- \rangle = 0$. In this case, pairs of the f-Exc have condensed, implying that
the interlayer  $U(1)_{\textrm{r}}$ is spontaneously broken down to $Z_2$. This leaves behind a mod-2 conservation law
for the exciton number. Since pairs of f-Exc are topologically equivalent to pairs of b-Exc,
this state can also be viewed as a state where \emph{pairs} of b-Exc have condensed.

Note that in both case (a) and (b), we can further consider various types of paired states for the $\psi_{\pm}$ fermions, e.g., whether they are weak or strong pairing superconductors~\cite{read2000}.
Wave functions for these interlayer coherent FQH states can be written as
$\Psi(\{x_i,\sigma_i\}) = \mbox{Pf}\left[  g_{\sigma _i \sigma_j}({\bf r}_i - \bf{r }_j)  \right] \prod_{i<j}(x_i - x_j)^2$, where $x_i$ is now the
complex coordinate of the $i$th electron including both layers and $\sigma_i = \pm$ is its layer index. $g_{\sigma_i \sigma_j}({\bf r}_i - {\bf r}_j)$ is the pair wave function.
For example, if we take $g_{\sigma_i \sigma_j}({\bf r}_i - {\bf r}_j) = \frac{\Delta_{\sigma \sigma'}}{x_i - x_j}$, this would
correspond to the case where $\langle \psi_\sigma(k) \psi_{\sigma'}(-k) \rangle = \Delta_{\sigma \sigma'} (k_x + i k_y)$.

(4) \it Pfaffian FQH states with localized excitons. \rm
Finally, we can consider a state where $\psi_+$ is paired, while the $\psi_-$ fermions form a density wave state, or, in the presence
of disorder, are localized. This is the state which, in the language of excitons used earlier, corresponds to a Pfaffian FQH state
in one layer with some density of localized excitons. As explained above, this disordered state is not a sharply distinct phase
from the $Z_2$ fractionalized exciton metal, but rather a different regime of the same phase.
The topological order of such a state is simply that of the Pfaffian FQH state, regardless of whether the b-Exc or f-Exc are lower in energy.

\paragraph{Exact diagonalization study of the Pfaffian's exciton energies.} While we have enumerated a variety of consistent possibilities,  it is a matter of microscopic energetics which will actually occur. A comprehensive numerical investigation is presented elsewhere,[PRB] but here we address the most important question: does the b-Exc, or f-Exc, have lower energy? We answer this question using exact diagonalization of the Coulomb Hamiltonian on a sphere, keeping both an $N=0$ and $N=1$ LL. \footnote{Note the two LLs have a different number of orbitals due to the ``shift"~\cite{wen1992shift}: the number of orbitals in $N=0$ LL is $N_{\textrm{orb}, 0} = N_{\phi} + 1$, where $N_\phi$ is the number of flux quanta, while in $N=1$ LL the number of orbitals is $N_{\textrm{orb}, 1} = N_{\phi} + 3$. This difference is important for accurate finite-size calculations.}

\begin{figure}[t]
\begin{center}
{\includegraphics[width=0.99\linewidth]{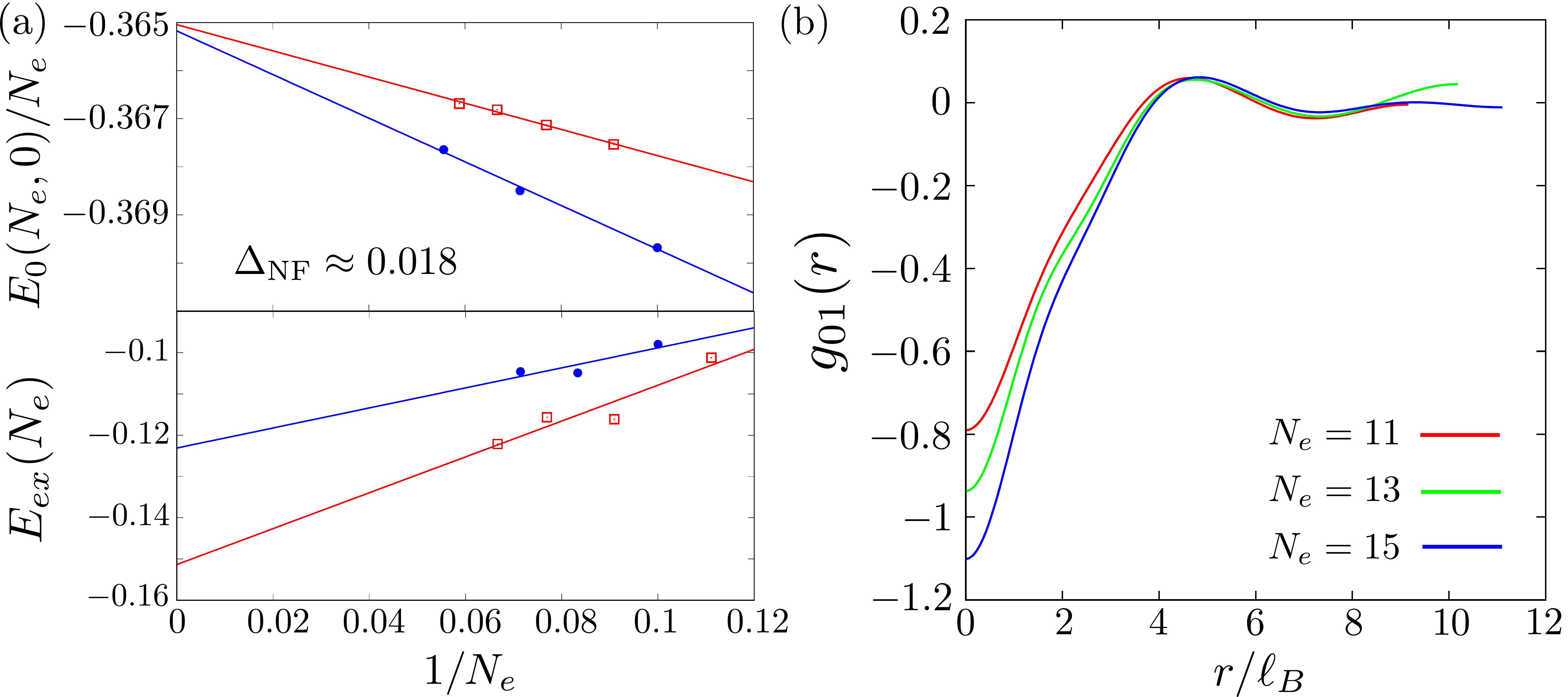}}
\caption{\label{fig:excitonED} (\textbf{a}) Top panel shows the energy per electron $E(N_1 = N_e, N_0 = 0) / N_e \to e_0$ when all electrons are in the $N=1$ layer. 
The  odd-even effect~\cite{Park2010} confirms that the Pfaffian ground state occurs for $N_e$-even, while $N_e$-odd corresponds to the $\nf$ excited state. Extrapolating the energy difference in $1 / N_e$ we obtain the neutral fermion gap $\Enf \sim 0.018$. In the bottom panel, we transfer one electron from the $N=1$ to the $N=0$ layer and measure the energy relative to the vacuum, $E_{ex}(N_e) = E(N_e - 1, 1) - e_0 N_e$. There is again an odd-even effect, but reversed: the bosonic exciton (blue, $N_e= 2 m$) is considerably higher in energy than the fermionic exciton (red, $N_e = 2 m + 1$).
(\textbf{b}) The f-Exc pair correlation function between the $N= 0, 1$ layers,  $g_{01}(r)$,  shows that the electron and hole bind together into an exciton of size $\sim 4 \ell_B$. }
\end{center}
\end{figure}

To explain the results in Fig.~\ref{fig:excitonED} we must recall some facts about the Pfaffian state on a sphere.
The Pfaffian ground state occurs when the number of electrons $N_e$ and the number of flux quanta  $N_{\phi}$ satisfies $N_{\phi} = 2 N_e - 5$.  When $N_e$ is even, the sphere has a unique, gapped ground state.  In the top panel of Fig.~\ref{fig:excitonED}(a), we show the energy per electron $E(N_1 = N_e, N_0 = 0) / N_e $ when all electrons are in the $N=1$ layer. Calculations are done for the Coulomb interaction with energies expressed in units of $e^2 / \epsilon \ell_B$.  Using standard finite-size corrections~\cite{Morf98,Morf02} and linear extrapolation in $1/N_e$  for $N_e$-even, we find the thermodynamic vacuum energy per particle of the Pfaffian to be $e_0 \approx -0.365$. 
However, when $N_e$ is odd, there is a dispersing band of low energy states~\cite{BondersonNeutralGap, MollerNeutralGap}. This can be understood by appealing to the ``superconducting'' nature of Pfaffian phase \cite{read2000}: when the number of CFs is odd, one CF must remain as an unpaired BdG quasiparticle, which is precisely the neutral fermion $\nf$ excitation.
By measuring the ground state energy differences $E(N_e) - e_0 N_e$, where $N_e$ is odd and $e_0$ is the energy per electron in the thermodynamic limit (top panel in Fig.~\ref{fig:excitonED}a), we estimate neutral fermion gap $\Enf \sim  0.018$, in line with earlier studies~\cite{BondersonNeutralGap, MollerNeutralGap}.

A similar method can be used to measure the energy difference between the f-Exc and b-Exc, see bottom panel of Fig.~\ref{fig:excitonED}(a).  Let $E(N_{1}, N_{0})$ be the ground state energy for $N_e = N_{1} + N_0$ electrons in the $N = 1, 0$ levels respectively, keeping fixed  the number of flux $N_{\phi} = 2 N_e - 5$.  The b-Exc occurs when $N_0 = 1$ and $N_e$ is even; in contrast, the f-Exc occurs for $N_0 = 1$ and $N_e$ is odd.
We define the exciton energies by subtracting off the Pfaffian's extrapolated vacuum energy $e_0$~\cite{Morf98,Morf02},
\begin{align}
E_{ex}(N_e) = E(N_e -1, 1) - e_0 N_e.
\end{align}

The exciton energy  $E_{ex}(N_e)$  also shows an odd-even  effect, Fig.~\ref{fig:excitonED}(a), but in contrast to the vacuum,  \emph{odd} $N_e$ (the f-Exc) is now lower in energy by about $\Eb - \Ef \gtrsim 0.02$.
Note that since the b-Exc can decay into an f-Exc and a $\nf$, we do not expect to see a difference much greater than $\Enf \approx 0.018$, though the energy of the \emph{metastable} b-Exc may be much larger.

To verify that the electron and hole are  forming a tightly bound exciton, we examine the inter-layer pair correlation function  in the f-Exc sector,  $g_{01}(r) \equiv A \langle\langle  \hat{n}_0(r) (\hat{n}_1(0) -  \bar{n}_1  ) \rangle \rangle$, where $\bar{n}_1=N_e/A$ is the average density in the Pfaffian ground state and $A$ is the area of the sphere. The f-Exc carries angular momentum $L=3/2$, so the double brackets denote an average over the $L$-multiplet.   We subtract $\bar{n}_1$ so that $- \int d^2 r g_{01}(r) = 1$ can interpreted as the probability for the electron and hole to be at distance $r$. As we see in Fig.~\ref{fig:excitonED}(b), they indeed bind together into an exciton of size $\sim 4 \ell_B$.

In summary, exact diagonalization of the Coulomb Hamiltonian  shows that as charge is transferred between layers the electrons and holes form tightly bound excitons, and the non-trivial f-Exc is the lowest energy exciton. At dilute exciton densities this ``single particle''  energy will dominate over interactions,  indicating that a fermionic exciton metal is more likely than a bosonic condensate.

A number of experimental signatures could be used to distinguish these scenarios:

 \it Counterflow\rm --
Counterflow transport is a clear way to distinguish between   localized Bose/Fermi excitons, interlayer coherent FQH states, and the exciton metal. Assuming the ability to independently contact the two layers, one can measure the counterflow
conductivity: $j_\textrm{r}= \sigma_\textrm{r} E_\textrm{r}$, where $j_\textrm{r} = j_+ - j_-$ is the relative current and
$E_\textrm{r} = E_+ - E_-$ is the difference in electric field between the two layers. When $\langle \psi_+ \psi_+ \rangle \neq 0$,
$j_\textrm{r}$ is simply the current of the $\psi_-$ fermions. The DC ``counterflow conductivity''
$\sigma_\textrm{r}$ will thus be zero, finite, or infinite, depending on whether the b-Exc have Bose condensed, the
f-Exc have formed a Fermi sea (with temperature $T$  greater than the localization cross-over scale), or the excitons have localized.
A \it dissipative \rm counterflow conductivity, in an incompressible FQH insulator, is a striking property of the exciton metal state.

\it Polarizability\rm-- The polarizability is defined 
as $\lim_{\omega \rightarrow 0, q \rightarrow 0} \langle p(q, \omega) p(-q,-\omega) \rangle$,
where $p(x, t) = n_+(x,t) - n_-(x,t)$ is the difference in density between the two components.
All states considered above have finite polarizability. When the excitons are localized by disorder in either the bosonic or fermionic case, the polarizability is set by the disorder strength; in the Bose exciton condensate state it is set by the superfluid density, and in the exciton Fermi sea it is set by the density of states at the Fermi surface.
The latter can be understood within the field theory presented above: if $\langle \psi_+ \psi_+ \rangle \neq 0$, then $\psi_-$ is a f-Exc, $p \sim \psi_-^\dagger \psi_- + const,$ and polarizability is simply the compressibility of the f-Exc state. The exciton Fermi sea can be distinguished the temperature dependence of the polarizability or by the application of a periodic potential: when the wave vector of the periodic potential becomes commensurate with $2k_F$, Bragg scattering induces an exciton band gap and modulates the polarizability.

\it Specific heat and thermal conductivity \rm--
Another characteristic distinguishing feature of the different exciton states appears in the specific heat and the thermal conductivity.
The thermal conductivity of the exciton metal will be linear in temperature: $\kappa \sim C_v v_F \ell \sim T$, where
$\ell$ is the mean free path of the excitons, $v_F$ is their Fermi velocity, and $C_v \sim T$ is the specific heat of the exciton Fermi surface.
Since such a state has zero electrical conductivity at zero temperature, this would imply an infinite violation of the Wiedemann-Franz law.
In contrast, the thermal conductivity of the exciton localized state $\kappa \rightarrow 0$ at zero temperature, although the
specific heat is still expected to be linear in $T$ in this phase.

\paragraph{$(N_+, N_-) = (0,0)$: (331) fractional exciton metal.}  Finally, we briefly mention an alternative platform for an exciton metal. In QH bilayers with $d/l_B > 1$ at filling $(\nu_+, \nu_-) = (1/4,1/4)$ the bilayer can form a $331$ state.
This state has been observed when both components partially fill the $N_{\pm} = 0$ LL,\cite{shayegan1999} though it may also happen more generally.
On the other hand, when $(\nu_+,\nu_-) = (1/2,0)$ or $(0,1/2)$, the system will form a CFL state.  What is the fate of the system in the intermediate regime $(\nu_+, \nu_-) = (1/4 + \delta, 1/4 - \delta)$? The  $331$  state also possesses an f-Exc,
which contains charge $e/2$ and $-e/2$ in the two layers, respectively 
Note that this is quite distinct from the scenario considered earlier, where the f-Exc in the Pfaffian state contained charge $e$ and $-e$ in the two layers.
Since the b-Exc has charge $e$ and $-e$ while the f-Exc has charge $e/2$ and $-e/2$, we expect that the Coulomb repulsion
would cause the b-Exc to be unstable to decaying into two f-Exc's. As $\delta$ is tuned away from zero, the finite density of f-Exc's
can form a Fermi sea. In terms of the effective theory, the $331$ can be described by writing $c_{\pm} = b_{\pm} \psi$, and assuming a
mean-field state where $b_{\pm}$ each form $\nu = 1/2$ bosonic Laughlin states in each layer, and $\psi$
forms a $\nu = 1$ IQH state. The exciton in this picture can be thought of as a pair of $\pm e/2$ quasiparticles of the bosonic Laughlin states in each layer.
As in the case of the Pfaffian exciton metal, the $331$ state also possesses charge $\pm e/4$ quasiparticles, which the f-Exc's
see as sources of an effective $\pi$-flux.

In summary, we have shown that bilayer QH systems at $\nu_{\textrm{T}} = \frac{1}{2}$ can support a variety of fractionalized exciton phases due to the interplay of U(1)$_{\textrm{r}}$ symmetry and even-denominator fractionalization.
We find numerical evidence that the lowest-energy exciton in these systems is a fermion, suggesting that the Z$_2$-fractionalized ``topological exciton metal'' is a strong possibility.
An even-denominator states at finite layer polarization has already been observed in BLG,\cite{zibrov2016}  so we hope future experiments can look for the dramatic transport signatures of this phase.

\it Acknowledgements\rm-- We thank Roger Mong for discussions.  AFY acknowledges the support of the Army Research Office under grant W911NF-16-1-0482.
MB is supported by startup funds from the University of Maryland and NSF-JQI-PFC. This research was supported in part by the National Science Foundation under Grant No. NSF PHY-1125915 (KITP). Z.P. acknowledges support by EPSRC grant EP/P009409/1. Statement of compliance with EPSRC policy framework on research data: This publication is theoretical work that does not require supporting research data.


\begin{thebibliography}{35}
\expandafter\ifx\csname natexlab\endcsname\relax\def\natexlab#1{#1}\fi
\expandafter\ifx\csname bibnamefont\endcsname\relax
  \def\bibnamefont#1{#1}\fi
\expandafter\ifx\csname bibfnamefont\endcsname\relax
  \def\bibfnamefont#1{#1}\fi
\expandafter\ifx\csname citenamefont\endcsname\relax
  \def\citenamefont#1{#1}\fi
\expandafter\ifx\csname url\endcsname\relax
  \def\url#1{\texttt{#1}}\fi
\expandafter\ifx\csname urlprefix\endcsname\relax\def\urlprefix{URL }\fi
\providecommand{\bibinfo}[2]{#2}
\providecommand{\eprint}[2][]{\url{#2}}

\bibitem[{\citenamefont{Girvin and MacDonald}(2007)}]{girvinmacdonald}
\bibinfo{author}{\bibfnamefont{S.~M.} \bibnamefont{Girvin}} \bibnamefont{and}
  \bibinfo{author}{\bibfnamefont{A.~H.} \bibnamefont{MacDonald}},
  \emph{\bibinfo{title}{Multicomponent Quantum Hall Systems: The Sum of Their
  Parts and More}} (\bibinfo{publisher}{Wiley-VCH Verlag GmbH},
  \bibinfo{year}{2007}), pp. \bibinfo{pages}{161--224}, ISBN
  \bibinfo{isbn}{9783527617258},
  \urlprefix\url{http://dx.doi.org/10.1002/9783527617258.ch5}.

\bibitem[{\citenamefont{Eisenstein}(2014)}]{eisenstein2014}
\bibinfo{author}{\bibfnamefont{J.}~\bibnamefont{Eisenstein}},
  \bibinfo{journal}{Annu. Rev. Condens. Matter Phys.}
  \textbf{\bibinfo{volume}{5}}, \bibinfo{pages}{159} (\bibinfo{year}{2014}).

\bibitem[{\citenamefont{Shayegan}(1999)}]{shayegan1999}
\bibinfo{author}{\bibfnamefont{E.}~\bibnamefont{Shayegan}}, in
  \emph{\bibinfo{booktitle}{1998 Les Houches Summer School, Session LXIX,
  Topological Aspects of Low Dimensional Systems, NATO Advanced Study
  Institute}}, edited by
  \bibinfo{editor}{\bibfnamefont{A.}~\bibnamefont{Comtet}},
  \bibinfo{editor}{\bibfnamefont{T.}~\bibnamefont{Jolicoeur}},
  \bibinfo{editor}{\bibfnamefont{S.}~\bibnamefont{Ouvry}}, \bibnamefont{and}
  \bibinfo{editor}{\bibfnamefont{F.}~\bibnamefont{David}}
  (\bibinfo{publisher}{Springer-Verlag}, \bibinfo{address}{Singapore},
  \bibinfo{year}{1999}), pp. \bibinfo{pages}{1--51}.

\bibitem[{\citenamefont{Eisenstein et~al.}(1989)\citenamefont{Eisenstein,
  Stormer, Pfeiffer, and West}}]{eisensteintransition1989}
\bibinfo{author}{\bibfnamefont{J.~P.} \bibnamefont{Eisenstein}},
  \bibinfo{author}{\bibfnamefont{H.~L.} \bibnamefont{Stormer}},
  \bibinfo{author}{\bibfnamefont{L.}~\bibnamefont{Pfeiffer}}, \bibnamefont{and}
  \bibinfo{author}{\bibfnamefont{K.~W.} \bibnamefont{West}},
  \bibinfo{journal}{Phys. Rev. Lett.} \textbf{\bibinfo{volume}{62}},
  \bibinfo{pages}{1540} (\bibinfo{year}{1989}),
  \urlprefix\url{http://link.aps.org/doi/10.1103/PhysRevLett.62.1540}.

\bibitem[{\citenamefont{Bishop et~al.}(2007)\citenamefont{Bishop, Padmanabhan,
  Vakili, Shkolnikov, De~Poortere, and Shayegan}}]{bishop2007}
\bibinfo{author}{\bibfnamefont{N.~C.} \bibnamefont{Bishop}},
  \bibinfo{author}{\bibfnamefont{M.}~\bibnamefont{Padmanabhan}},
  \bibinfo{author}{\bibfnamefont{K.}~\bibnamefont{Vakili}},
  \bibinfo{author}{\bibfnamefont{Y.~P.} \bibnamefont{Shkolnikov}},
  \bibinfo{author}{\bibfnamefont{E.~P.} \bibnamefont{De~Poortere}},
  \bibnamefont{and} \bibinfo{author}{\bibfnamefont{M.}~\bibnamefont{Shayegan}},
  \bibinfo{journal}{Phys. Rev. Lett.} \textbf{\bibinfo{volume}{98}},
  \bibinfo{pages}{266404} (\bibinfo{year}{2007}).

\bibitem[{\citenamefont{Feldman et~al.}(2013)\citenamefont{Feldman, Levin,
  Krauss, Abanin, Halperin, Smet, and Yacoby}}]{feldmantransition2013}
\bibinfo{author}{\bibfnamefont{B.~E.} \bibnamefont{Feldman}},
  \bibinfo{author}{\bibfnamefont{A.~J.} \bibnamefont{Levin}},
  \bibinfo{author}{\bibfnamefont{B.}~\bibnamefont{Krauss}},
  \bibinfo{author}{\bibfnamefont{D.~A.} \bibnamefont{Abanin}},
  \bibinfo{author}{\bibfnamefont{B.~I.} \bibnamefont{Halperin}},
  \bibinfo{author}{\bibfnamefont{J.~H.} \bibnamefont{Smet}}, \bibnamefont{and}
  \bibinfo{author}{\bibfnamefont{A.}~\bibnamefont{Yacoby}},
  \bibinfo{journal}{Phys. Rev. Lett.} \textbf{\bibinfo{volume}{111}},
  \bibinfo{pages}{076802} (\bibinfo{year}{2013}).

\bibitem[{\citenamefont{Weitz et~al.}(2010)\citenamefont{Weitz, Allen, Feldman,
  Martin, and Yacoby}}]{weitz_broken-symmetry_2010}
\bibinfo{author}{\bibfnamefont{R.~T.} \bibnamefont{Weitz}},
  \bibinfo{author}{\bibfnamefont{M.~T.} \bibnamefont{Allen}},
  \bibinfo{author}{\bibfnamefont{B.~E.} \bibnamefont{Feldman}},
  \bibinfo{author}{\bibfnamefont{J.}~\bibnamefont{Martin}}, \bibnamefont{and}
  \bibinfo{author}{\bibfnamefont{A.}~\bibnamefont{Yacoby}},
  \bibinfo{journal}{Science} \textbf{\bibinfo{volume}{330}},
  \bibinfo{pages}{812} (\bibinfo{year}{2010}),
  \urlprefix\url{http://www.sciencemag.org/content/330/6005/812.abstract}.

\bibitem[{\citenamefont{Falson et~al.}(2015)\citenamefont{Falson, Maryenko,
  Friess, Zhang, Kozuka, Tsukazaki, Smet, and
  Kawasaki}}]{falson_even-denominator_2015}
\bibinfo{author}{\bibfnamefont{J.}~\bibnamefont{Falson}},
  \bibinfo{author}{\bibfnamefont{D.}~\bibnamefont{Maryenko}},
  \bibinfo{author}{\bibfnamefont{B.}~\bibnamefont{Friess}},
  \bibinfo{author}{\bibfnamefont{D.}~\bibnamefont{Zhang}},
  \bibinfo{author}{\bibfnamefont{Y.}~\bibnamefont{Kozuka}},
  \bibinfo{author}{\bibfnamefont{A.}~\bibnamefont{Tsukazaki}},
  \bibinfo{author}{\bibfnamefont{J.~H.} \bibnamefont{Smet}}, \bibnamefont{and}
  \bibinfo{author}{\bibfnamefont{M.}~\bibnamefont{Kawasaki}},
  \bibinfo{journal}{Nature Physics} \textbf{\bibinfo{volume}{11}},
  \bibinfo{pages}{347} (\bibinfo{year}{2015}), ISSN \bibinfo{issn}{1745-2473},
  \urlprefix\url{http://www.nature.com/nphys/journal/v11/n4/full/nphys3259.html}.

\bibitem[{\citenamefont{Eisenstein et~al.}(1992)\citenamefont{Eisenstein,
  Boebinger, Pfeiffer, West, and He}}]{eisenstein1992a}
\bibinfo{author}{\bibfnamefont{J.~P.} \bibnamefont{Eisenstein}},
  \bibinfo{author}{\bibfnamefont{G.~S.} \bibnamefont{Boebinger}},
  \bibinfo{author}{\bibfnamefont{L.~N.} \bibnamefont{Pfeiffer}},
  \bibinfo{author}{\bibfnamefont{K.~W.} \bibnamefont{West}}, \bibnamefont{and}
  \bibinfo{author}{\bibfnamefont{S.}~\bibnamefont{He}}, \bibinfo{journal}{Phys.
  Rev. Lett.} \textbf{\bibinfo{volume}{68}}, \bibinfo{pages}{1383}
  (\bibinfo{year}{1992}).

\bibitem[{\citenamefont{Suen et~al.}(1992)\citenamefont{Suen, Engel, Santos,
  Shayegan, and Tsui}}]{suen1992}
\bibinfo{author}{\bibfnamefont{Y.~W.} \bibnamefont{Suen}},
  \bibinfo{author}{\bibfnamefont{L.~W.} \bibnamefont{Engel}},
  \bibinfo{author}{\bibfnamefont{M.~B.} \bibnamefont{Santos}},
  \bibinfo{author}{\bibfnamefont{M.}~\bibnamefont{Shayegan}}, \bibnamefont{and}
  \bibinfo{author}{\bibfnamefont{D.~C.} \bibnamefont{Tsui}},
  \bibinfo{journal}{Phys. Rev. Lett.} \textbf{\bibinfo{volume}{68}},
  \bibinfo{pages}{1379} (\bibinfo{year}{1992}),
  \urlprefix\url{http://link.aps.org/doi/10.1103/PhysRevLett.68.1379}.

\bibitem[{\citenamefont{Suen et~al.}(1994)\citenamefont{Suen, Manoharan, Ying,
  Santos, and Shayegan}}]{suen1994}
\bibinfo{author}{\bibfnamefont{Y.~W.} \bibnamefont{Suen}},
  \bibinfo{author}{\bibfnamefont{H.~C.} \bibnamefont{Manoharan}},
  \bibinfo{author}{\bibfnamefont{X.}~\bibnamefont{Ying}},
  \bibinfo{author}{\bibfnamefont{M.~B.} \bibnamefont{Santos}},
  \bibnamefont{and} \bibinfo{author}{\bibfnamefont{M.}~\bibnamefont{Shayegan}},
  \bibinfo{journal}{Phys. Rev. Lett.} \textbf{\bibinfo{volume}{72}},
  \bibinfo{pages}{3405} (\bibinfo{year}{1994}).

\bibitem[{\citenamefont{Liu et~al.}(2011)\citenamefont{Liu, Shabani, Kamburov,
  Shayegan, Pfeiffer, West, and Baldwin}}]{liu2011}
\bibinfo{author}{\bibfnamefont{Y.}~\bibnamefont{Liu}},
  \bibinfo{author}{\bibfnamefont{J.}~\bibnamefont{Shabani}},
  \bibinfo{author}{\bibfnamefont{D.}~\bibnamefont{Kamburov}},
  \bibinfo{author}{\bibfnamefont{M.}~\bibnamefont{Shayegan}},
  \bibinfo{author}{\bibfnamefont{L.~N.} \bibnamefont{Pfeiffer}},
  \bibinfo{author}{\bibfnamefont{K.~W.} \bibnamefont{West}}, \bibnamefont{and}
  \bibinfo{author}{\bibfnamefont{K.~W.} \bibnamefont{Baldwin}},
  \bibinfo{journal}{Phys. Rev. Lett.} \textbf{\bibinfo{volume}{107}},
  \bibinfo{pages}{266802} (\bibinfo{year}{2011}).

\bibitem[{\citenamefont{Liu et~al.}(2014{\natexlab{a}})\citenamefont{Liu,
  Hasdemir, Kamburov, Graninger, Shayegan, Pfeiffer, West, Baldwin, and
  Winkler}}]{liu_even-denominator_2014}
\bibinfo{author}{\bibfnamefont{Y.}~\bibnamefont{Liu}},
  \bibinfo{author}{\bibfnamefont{S.}~\bibnamefont{Hasdemir}},
  \bibinfo{author}{\bibfnamefont{D.}~\bibnamefont{Kamburov}},
  \bibinfo{author}{\bibfnamefont{A.~L.} \bibnamefont{Graninger}},
  \bibinfo{author}{\bibfnamefont{M.}~\bibnamefont{Shayegan}},
  \bibinfo{author}{\bibfnamefont{L.~N.} \bibnamefont{Pfeiffer}},
  \bibinfo{author}{\bibfnamefont{K.~W.} \bibnamefont{West}},
  \bibinfo{author}{\bibfnamefont{K.~W.} \bibnamefont{Baldwin}},
  \bibnamefont{and} \bibinfo{author}{\bibfnamefont{R.}~\bibnamefont{Winkler}},
  \bibinfo{journal}{Physical Review B} \textbf{\bibinfo{volume}{89}},
  \bibinfo{pages}{165313} (\bibinfo{year}{2014}{\natexlab{a}}),
  \urlprefix\url{http://link.aps.org/doi/10.1103/PhysRevB.89.165313}.

\bibitem[{\citenamefont{Liu et~al.}(2014{\natexlab{b}})\citenamefont{Liu,
  Graninger, Hasdemir, Shayegan, Pfeiffer, West, Baldwin, and
  Winkler}}]{liu2014prl}
\bibinfo{author}{\bibfnamefont{Y.}~\bibnamefont{Liu}},
  \bibinfo{author}{\bibfnamefont{A.~L.} \bibnamefont{Graninger}},
  \bibinfo{author}{\bibfnamefont{S.}~\bibnamefont{Hasdemir}},
  \bibinfo{author}{\bibfnamefont{M.}~\bibnamefont{Shayegan}},
  \bibinfo{author}{\bibfnamefont{L.~N.} \bibnamefont{Pfeiffer}},
  \bibinfo{author}{\bibfnamefont{K.~W.} \bibnamefont{West}},
  \bibinfo{author}{\bibfnamefont{K.~W.} \bibnamefont{Baldwin}},
  \bibnamefont{and} \bibinfo{author}{\bibfnamefont{R.}~\bibnamefont{Winkler}},
  \bibinfo{journal}{Phys. Rev. Lett.} \textbf{\bibinfo{volume}{112}},
  \bibinfo{pages}{046804} (\bibinfo{year}{2014}{\natexlab{b}}).

\bibitem[{\citenamefont{Ki et~al.}(2014)\citenamefont{Ki, Fal’ko, Abanin, and
  Morpurgo}}]{ki2014}
\bibinfo{author}{\bibfnamefont{D.-K.} \bibnamefont{Ki}},
  \bibinfo{author}{\bibfnamefont{V.~I.} \bibnamefont{Fal’ko}},
  \bibinfo{author}{\bibfnamefont{D.~A.} \bibnamefont{Abanin}},
  \bibnamefont{and} \bibinfo{author}{\bibfnamefont{A.~F.}
  \bibnamefont{Morpurgo}}, \bibinfo{journal}{Nano Letters}
  \textbf{\bibinfo{volume}{14}}, \bibinfo{pages}{2135} (\bibinfo{year}{2014}),
  \bibinfo{note}{pMID: 24611523}, \eprint{http://dx.doi.org/10.1021/nl5003922}.

\bibitem[{\citenamefont{Zibrov et~al.}(2016)\citenamefont{Zibrov, Kometter,
  Taniguchi, Watanabe, Zaletel, and Young}}]{zibrov2016}
\bibinfo{author}{\bibfnamefont{A.}~\bibnamefont{Zibrov}},
  \bibinfo{author}{\bibfnamefont{C.}~\bibnamefont{Kometter}},
  \bibinfo{author}{\bibfnamefont{T.}~\bibnamefont{Taniguchi}},
  \bibinfo{author}{\bibfnamefont{K.}~\bibnamefont{Watanabe}},
  \bibinfo{author}{\bibfnamefont{M.}~\bibnamefont{Zaletel}}, \bibnamefont{and}
  \bibinfo{author}{\bibfnamefont{A.}~\bibnamefont{Young}}, \bibinfo{journal}{In
  preparation}  (\bibinfo{year}{2016}).

\bibitem[{\citenamefont{Willett et~al.}(1987)\citenamefont{Willett, Eisenstein,
  St\"ormer, Tsui, Gossard, and English}}]{willett1987}
\bibinfo{author}{\bibfnamefont{R.}~\bibnamefont{Willett}},
  \bibinfo{author}{\bibfnamefont{J.~P.} \bibnamefont{Eisenstein}},
  \bibinfo{author}{\bibfnamefont{H.~L.} \bibnamefont{St\"ormer}},
  \bibinfo{author}{\bibfnamefont{D.~C.} \bibnamefont{Tsui}},
  \bibinfo{author}{\bibfnamefont{A.~C.} \bibnamefont{Gossard}},
  \bibnamefont{and} \bibinfo{author}{\bibfnamefont{J.~H.}
  \bibnamefont{English}}, \bibinfo{journal}{Phys. Rev. Lett.}
  \textbf{\bibinfo{volume}{59}}, \bibinfo{pages}{1776} (\bibinfo{year}{1987}).

\bibitem[{\citenamefont{Morf}(1998)}]{Morf98}
\bibinfo{author}{\bibfnamefont{R.~H.} \bibnamefont{Morf}},
  \bibinfo{journal}{Phys. Rev. Lett.} \textbf{\bibinfo{volume}{80}},
  \bibinfo{pages}{1505} (\bibinfo{year}{1998}),
  \urlprefix\url{https://link.aps.org/doi/10.1103/PhysRevLett.80.1505}.

\bibitem[{\citenamefont{Banerjee et~al.}(2017)\citenamefont{Banerjee, Heiblum,
  Umansky, Feldman, Oreg, and Stern}}]{banerjee2017observation}
\bibinfo{author}{\bibfnamefont{M.}~\bibnamefont{Banerjee}},
  \bibinfo{author}{\bibfnamefont{M.}~\bibnamefont{Heiblum}},
  \bibinfo{author}{\bibfnamefont{V.}~\bibnamefont{Umansky}},
  \bibinfo{author}{\bibfnamefont{D.~E.} \bibnamefont{Feldman}},
  \bibinfo{author}{\bibfnamefont{Y.}~\bibnamefont{Oreg}}, \bibnamefont{and}
  \bibinfo{author}{\bibfnamefont{A.}~\bibnamefont{Stern}},
  \bibinfo{journal}{arXiv:1710.00492}  (\bibinfo{year}{2017}).

\bibitem[{\citenamefont{Moore and Read}(1991)}]{moore1991}
\bibinfo{author}{\bibfnamefont{G.}~\bibnamefont{Moore}} \bibnamefont{and}
  \bibinfo{author}{\bibfnamefont{N.}~\bibnamefont{Read}},
  \bibinfo{journal}{Nuclear Physics B} \textbf{\bibinfo{volume}{360}},
  \bibinfo{pages}{362 } (\bibinfo{year}{1991}).

\bibitem[{\citenamefont{Willett}(1997)}]{willett1997}
\bibinfo{author}{\bibfnamefont{R.~L.} \bibnamefont{Willett}},
  \bibinfo{journal}{Advances in Physics} \textbf{\bibinfo{volume}{46}},
  \bibinfo{pages}{447} (\bibinfo{year}{1997}).

\bibitem[{\citenamefont{Halperin et~al.}(1993)\citenamefont{Halperin, Lee, and
  Read}}]{halperin1993}
\bibinfo{author}{\bibfnamefont{B.~I.} \bibnamefont{Halperin}},
  \bibinfo{author}{\bibfnamefont{P.~A.} \bibnamefont{Lee}}, \bibnamefont{and}
  \bibinfo{author}{\bibfnamefont{N.}~\bibnamefont{Read}},
  \bibinfo{journal}{Phys. Rev. B} \textbf{\bibinfo{volume}{47}},
  \bibinfo{pages}{7312} (\bibinfo{year}{1993}).

\bibitem[{\citenamefont{K.Jain}(2007)}]{jainCF}
\bibinfo{author}{\bibfnamefont{J.}~\bibnamefont{K.Jain}},
  \emph{\bibinfo{title}{Composite Fermions}} (\bibinfo{publisher}{Cambridge
  University Press}, \bibinfo{year}{2007}).

\bibitem[{\citenamefont{Jamei et~al.}(2005)\citenamefont{Jamei, Kivelson, and
  Spivak}}]{Jamei}
\bibinfo{author}{\bibfnamefont{R.}~\bibnamefont{Jamei}},
  \bibinfo{author}{\bibfnamefont{S.}~\bibnamefont{Kivelson}}, \bibnamefont{and}
  \bibinfo{author}{\bibfnamefont{B.}~\bibnamefont{Spivak}},
  \bibinfo{journal}{Phys. Rev. Lett.} \textbf{\bibinfo{volume}{94}},
  \bibinfo{pages}{056805} (\bibinfo{year}{2005}),
  \urlprefix\url{https://link.aps.org/doi/10.1103/PhysRevLett.94.056805}.

\bibitem[{\citenamefont{Fisher et~al.}(1989)\citenamefont{Fisher, Weichman,
  Grinstein, and Fisher}}]{FisherBoson}
\bibinfo{author}{\bibfnamefont{M.~P.~A.} \bibnamefont{Fisher}},
  \bibinfo{author}{\bibfnamefont{P.~B.} \bibnamefont{Weichman}},
  \bibinfo{author}{\bibfnamefont{G.}~\bibnamefont{Grinstein}},
  \bibnamefont{and} \bibinfo{author}{\bibfnamefont{D.~S.}
  \bibnamefont{Fisher}}, \bibinfo{journal}{Phys. Rev. B}
  \textbf{\bibinfo{volume}{40}}, \bibinfo{pages}{546} (\bibinfo{year}{1989}),
  \urlprefix\url{https://link.aps.org/doi/10.1103/PhysRevB.40.546}.

\bibitem[{\citenamefont{Lee and Ramakrishnan}(1985)}]{LeeReview}
\bibinfo{author}{\bibfnamefont{P.~A.} \bibnamefont{Lee}} \bibnamefont{and}
  \bibinfo{author}{\bibfnamefont{T.~V.} \bibnamefont{Ramakrishnan}},
  \bibinfo{journal}{Rev. Mod. Phys.} \textbf{\bibinfo{volume}{57}},
  \bibinfo{pages}{287} (\bibinfo{year}{1985}),
  \urlprefix\url{https://link.aps.org/doi/10.1103/RevModPhys.57.287}.

\bibitem[{\citenamefont{Barkeshli and McGreevy}(2012)}]{barkeshli2012hlr}
\bibinfo{author}{\bibfnamefont{M.}~\bibnamefont{Barkeshli}} \bibnamefont{and}
  \bibinfo{author}{\bibfnamefont{J.}~\bibnamefont{McGreevy}},
  \bibinfo{journal}{Phys. Rev. B} \textbf{\bibinfo{volume}{86}},
  \bibinfo{pages}{075136} (\bibinfo{year}{2012}).

\bibitem[{\citenamefont{Wen}(2004)}]{wen04}
\bibinfo{author}{\bibfnamefont{X.-G.} \bibnamefont{Wen}},
  \emph{\bibinfo{title}{Quantum Field Theory of Many-Body Systems}}
  (\bibinfo{publisher}{Oxford Univ. Press}, \bibinfo{address}{Oxford},
  \bibinfo{year}{2004}).

\bibitem[{\citenamefont{Read and Green}(2000)}]{read2000}
\bibinfo{author}{\bibfnamefont{N.}~\bibnamefont{Read}} \bibnamefont{and}
  \bibinfo{author}{\bibfnamefont{D.}~\bibnamefont{Green}},
  \bibinfo{journal}{Phys. Rev. B} \textbf{\bibinfo{volume}{61}},
  \bibinfo{pages}{10267} (\bibinfo{year}{2000}).

\bibitem[{\citenamefont{Nandkishore et~al.}(2012)\citenamefont{Nandkishore,
  Metlitski, and Senthil}}]{Nandkishore2012}
\bibinfo{author}{\bibfnamefont{R.}~\bibnamefont{Nandkishore}},
  \bibinfo{author}{\bibfnamefont{M.~A.} \bibnamefont{Metlitski}},
  \bibnamefont{and} \bibinfo{author}{\bibfnamefont{T.}~\bibnamefont{Senthil}},
  \bibinfo{journal}{Phys. Rev. B} \textbf{\bibinfo{volume}{86}},
  \bibinfo{pages}{045128} (\bibinfo{year}{2012}),
  \urlprefix\url{http://link.aps.org/doi/10.1103/PhysRevB.86.045128}.

\bibitem[{\citenamefont{Lu et~al.}(2010)\citenamefont{Lu, Das~Sarma, and
  Park}}]{Park2010}
\bibinfo{author}{\bibfnamefont{H.}~\bibnamefont{Lu}},
  \bibinfo{author}{\bibfnamefont{S.}~\bibnamefont{Das~Sarma}},
  \bibnamefont{and} \bibinfo{author}{\bibfnamefont{K.}~\bibnamefont{Park}},
  \bibinfo{journal}{Phys. Rev. B} \textbf{\bibinfo{volume}{82}},
  \bibinfo{pages}{201303} (\bibinfo{year}{2010}),
  \urlprefix\url{https://link.aps.org/doi/10.1103/PhysRevB.82.201303}.

\bibitem[{\citenamefont{Morf et~al.}(2002)\citenamefont{Morf, d'Ambrumenil, and
  Das~Sarma}}]{Morf02}
\bibinfo{author}{\bibfnamefont{R.~H.} \bibnamefont{Morf}},
  \bibinfo{author}{\bibfnamefont{N.}~\bibnamefont{d'Ambrumenil}},
  \bibnamefont{and}
  \bibinfo{author}{\bibfnamefont{S.}~\bibnamefont{Das~Sarma}},
  \bibinfo{journal}{Phys. Rev. B} \textbf{\bibinfo{volume}{66}},
  \bibinfo{pages}{075408} (\bibinfo{year}{2002}),
  \urlprefix\url{https://link.aps.org/doi/10.1103/PhysRevB.66.075408}.

\bibitem[{\citenamefont{Bonderson et~al.}(2011)\citenamefont{Bonderson,
  Feiguin, and Nayak}}]{BondersonNeutralGap}
\bibinfo{author}{\bibfnamefont{P.}~\bibnamefont{Bonderson}},
  \bibinfo{author}{\bibfnamefont{A.~E.} \bibnamefont{Feiguin}},
  \bibnamefont{and} \bibinfo{author}{\bibfnamefont{C.}~\bibnamefont{Nayak}},
  \bibinfo{journal}{Phys. Rev. Lett.} \textbf{\bibinfo{volume}{106}},
  \bibinfo{pages}{186802} (\bibinfo{year}{2011}),
  \urlprefix\url{http://link.aps.org/doi/10.1103/PhysRevLett.106.186802}.

\bibitem[{\citenamefont{M\"oller et~al.}(2011)\citenamefont{M\"oller, W\'ojs,
  and Cooper}}]{MollerNeutralGap}
\bibinfo{author}{\bibfnamefont{G.}~\bibnamefont{M\"oller}},
  \bibinfo{author}{\bibfnamefont{A.}~\bibnamefont{W\'ojs}}, \bibnamefont{and}
  \bibinfo{author}{\bibfnamefont{N.~R.} \bibnamefont{Cooper}},
  \bibinfo{journal}{Phys. Rev. Lett.} \textbf{\bibinfo{volume}{107}},
  \bibinfo{pages}{036803} (\bibinfo{year}{2011}),
  \urlprefix\url{http://link.aps.org/doi/10.1103/PhysRevLett.107.036803}.

\bibitem[{\citenamefont{Wen and Zee}(1992)}]{wen1992shift}
\bibinfo{author}{\bibfnamefont{X.~G.} \bibnamefont{Wen}} \bibnamefont{and}
  \bibinfo{author}{\bibfnamefont{A.}~\bibnamefont{Zee}},
  \bibinfo{journal}{Phys. Rev. Lett.} \textbf{\bibinfo{volume}{69}},
  \bibinfo{pages}{953} (\bibinfo{year}{1992}).

\end{thebibliography}

\end{document}